\documentclass[aps, prl, twocolumn,10pt,longbibliography]{revtex4-1}

\usepackage{amsmath, amssymb, amsfonts, amsthm, bm, mathrsfs, bbm}
\usepackage{graphicx}
\usepackage{epstopdf}
\usepackage{color}
\usepackage[usenames,dvipsnames]{xcolor}
\usepackage[citecolor=blue, colorlinks=true, urlcolor=blue, linkcolor=blue]{hyperref}

\newcommand{\eq}[1]{Eq.~(\ref{#1})}

\newcommand{\fig}[1]{Fig.\thinspace{}\ref{#1}}

\newcommand{\fc}[1]{({#1})}
\newcommand{\figc}[2]{Fig.\thinspace{}\ref{#1}\thinspace{}\fc{#2}}

\begin{document}

\title{Floquet prethermalization and regimes of heating in a periodically driven, interacting quantum system}

\author{Simon A. Weidinger}
\email[]{simon.weidinger@tum.de}

\author{Michael Knap}
\email[]{michael.knap@ph.tum.de}

\affiliation{Department of Physics, Walter Schottky Institute, and Institute for Advanced Study, Technical University of Munich, 85748 Garching, Germany}

\begin{abstract}

We study the regimes of heating in the periodically driven $O(N)$-model, which represents a generic model for interacting quantum many-body systems. By computing the absorbed energy with a non-equilibrium Keldysh Green's function approach, we establish three dynamical regimes: at short times a single-particle dominated regime, at intermediate times a stable Floquet prethermal regime in which the system ceases to absorb, and at parametrically late times a thermalizing regime. Our simulations suggest that in the thermalizing regime the absorbed energy grows algebraically in time with an the exponent that approaches the universal value of $1/2$, and is thus significantly slower than linear Joule heating. 
Our results demonstrate the parametric stability of prethermal states in a generic many-body system driven at frequencies that are comparable to its microscopic scales. This paves the way for realizing exotic quantum phases, such as time crystals or interacting topological phases, in the prethermal regime of interacting Floquet systems.

\end{abstract}

\date{\today}

\pacs{
}

\maketitle

Periodically driving quantum many-body systems often leads to exotic phenomena that are absent in their undriven counterparts. The unitary quantum evolution of a periodically driven system at times that are commensurate with the drive period $T$ is governed by the operator $\hat U(T)=\exp[-i \hat H_F T]$, which defines the Floquet Hamiltonian $\hat H_F$. The Floquet Hamiltonian $\hat H_F$ can be designed in such a way that it hosts novel and exotic phases of matter. Examples include, topologically non-trivial band structures realized by driving topologically trivial systems~\cite{Oka2009, Kitagawa2011, Lindner2010, Aidelsburger2013, Miyake2013, Jotzu2014, Aidelsburger2015}, and ergodic phases created by driving non-ergodic quantum systems~\cite{Ponte2015, Achilleas15, abanin_theory_2014, Kozarzewski_Distinctive_2016, Rehn_how_2016, gopalakrishnan_regimes_2016, bordia_periodically_2016}. Moreover, phases in periodically driven systems with no direct equilibrium analogue have been proposed~\cite{PSKhemani16, PSElse16, PS1Keyserlingk16, PS2Keyserlingk16, PSPotter16, FTCElse16, PS3Keyserlingk16, Else2016, YaoPotter16}, including Floquet time crystals which exhibit persistent macroscopic oscillations at integer multiples of the driving period~\cite{PSKhemani16, PS2Keyserlingk16, FTCElse16,  PS3Keyserlingk16, Else2016, YaoPotter16}. 

\begin{figure}
\includegraphics[width=.48\textwidth]{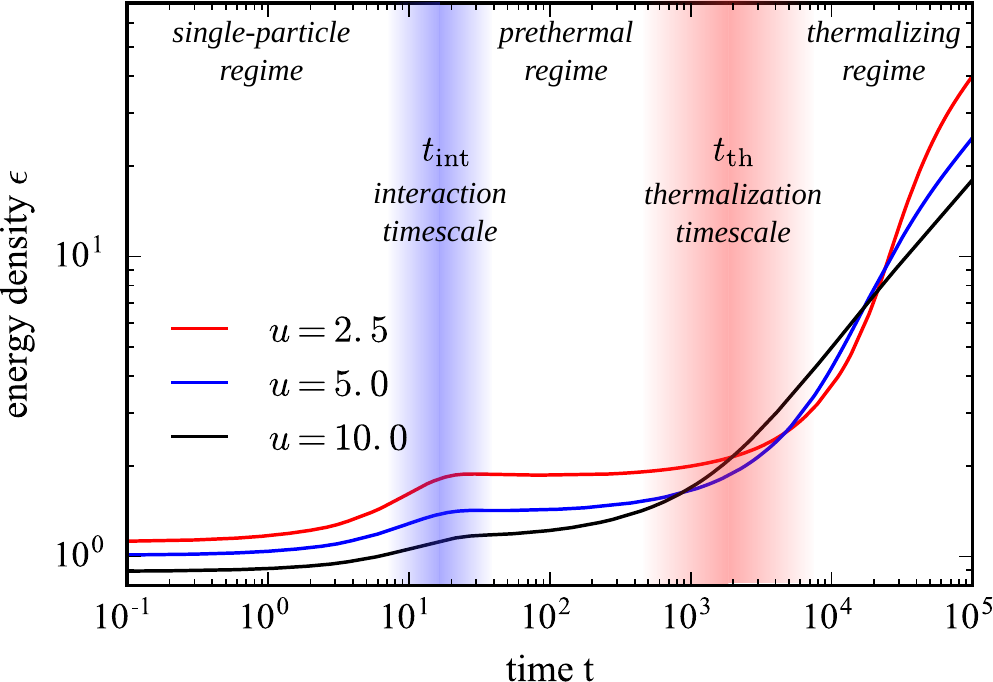}
\caption{\textbf{Time evolution of the energy density in the periodically driven $O(N)$ model.} The energy density $\epsilon(t)$ exhibits three distinct dynamical regimes: (I) At short times $t$, single-particle rearrangements lead to a fast increase of the energy density, up to times $t_\text{int}$ at which interactions become relevant. (II) At intermediate times, $t_\text{int} < t < t_\text{th}$, a stable Floquet prethermal regime occurs in which the interacting system ceases to absorb. (III) At late times, beyond the thermalization time scale $t>t_\text{th}$, interactions between a large number of generated quasi-particle excitations cause strong heating. In that regime, the energy density displays an algebraic growth, $\epsilon(t)\sim t^\alpha$, with an exponent that approaches $\alpha \sim 1/2$ for strong interactions. The data is shown for drive frequency $\Omega=2.3$ and for three different values of the interaction strength $u$, see legend.
}
\label{fig:timetraces}
\end{figure}

The eigenstate thermalization hypothesis (ETH) suggests that generic interacting many-body systems heat up to infinite temperature~\cite{DAlessio2014,Lazarides2014}, thus inhibiting the realization of such novel phases. A possible resolution is to stabilize the Floquet states by disorder such that the system becomes many-body localized and ETH does not apply~\cite{Ponte2015, Achilleas15, abanin_theory_2014, gopalakrishnan_regimes_2016}, as recently demonstrated experimentally~\cite{bordia_periodically_2016}. However, this restricts the variety of accessible phases. Another route would be to resort to driving frequencies much higher than all other microscopic scales~\cite{Abanin2015b, Mori2016}. But in that case $\hat{H}_\mathrm{F}$ becomes quasi-local and cannot possess any exotic phases. A more general approach, is to resort to a transient \emph{prethermal} regime~\cite{BukovGopalakrishnan2015, CanoviKollar2015, Chandran2016a, LindnerBerg2016}, which is characterized by the interaction time scale of the Floquet Hamiltonian required to realize exotic phenomena being much shorter than the heating timescale. It is therefore eminent to study the stability of such a Floquet prethermal regime in a general context.

In this work, we investigate the stability of the Floquet prethermal regime and the thermalization time scales in a generic interacting many-body system subject to periodic drive. In particular, we focus on the quantum $O(N)$-model with modulated mass. To this end, we employ the 2-particle irreducible (2PI) effective action approach on the closed Keldysh contour including corrections up to next-to-leading order (NLO) in $1/N$ which allow the system to thermalize. The $O(N)$-model is a generic and well established model for interacting many-body systems, both in condensed matter and cosmology~\cite{Cooper1997, Boyanovsky1996, Berges2002, Berges2007, SotiriadisCardy2010, Sciolla2013, ChandranNanduri2013, Smacchia2015, ChiocchettaTavora2015, Chandran2016a}. In particular, the presence of nontrivial interactions at NLO as well as the bosonic nature of excitations render the $O(N)$-model useful for studying heating of a driven many-body system to infinite temperature. Based on our numerical simulations, we find a parametrically large regime of Floquet prethermalization, even when the driving frequency is comparable to other microscopic scales of the undriven Hamiltonian so long as the interactions of the system are not too strong; \fig{fig:timetraces}. 

\begin{figure*}
\includegraphics[width=.98\textwidth]{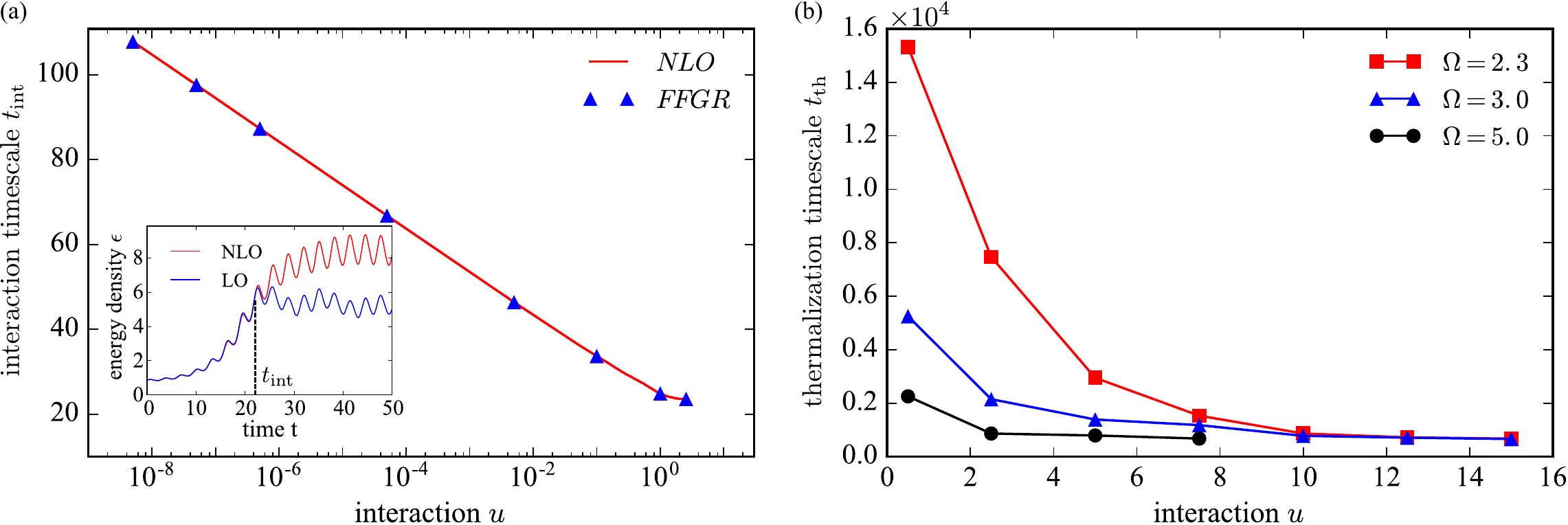}
\caption{\textbf{Parametric dependence of the interaction and thermalization time scales.} \fc{a} Interaction times scale $t_\mathrm{int}$ as a function of the interaction strength $u$. In the limit of weak interactions, $u \lesssim 1$, the interaction time $t_\text{int}$ scales logarithmically with $u$. This behavior can be analytically understood from a calculation based on Floquet Fermi's Golden Rule (FFGR), symbols, which perfectly agrees with the numerically evaluated interaction time scale, solid line. For $u \gtrsim1$, $t_\text{int}$ is not well defined, as interaction effects matter as soon as the drive is switched on. The inset illustrates the definition of $t_\mathrm{int}$ by comparing the time evolution of the energy density using leading order (LO) and next-to-leading order (NLO) approximations. At LO the heating stops at $t_\mathrm{int}$ while at NLO the system very slowly absorbs energy from the drive and enters the Floquet prethermalization regime. \fc{b} Thermalization timescale $t_\mathrm{th}$ as a function of the interaction strength $u$. The thermalization time scale $t_\mathrm{th}$ characterizes the crossover between the prethermal and the heating regime. It depends strongly on both the interactions $u$ and drive frequency $\Omega$.
}
\label{fig:tth}
\end{figure*}

\textit{\textbf{Model.---}}We study the quantum $O(N)$-model of $N$ real scalar fields $\Phi_a$, $a = 1,\dots, N$ with the action~\cite{MosheJustin-Zinn2003}
\begin{equation}
S= \int_{x, \mathcal{C}} \left[\frac{1}{2}(\partial_\mu\Phi_a)^2- \frac{1}{2} m^2(t) \Phi_a^2 - \frac{\lambda}{4! N} \left(\Phi_a \Phi_a\right)^2\right].
\label{eq:classicalS}
\end{equation}
We use the abbreviation $\int_{x, \mathcal{C}} = \int d^dx \int_\mathcal{C} dt$, where the time integration runs over the closed-time Keldysh contour $\mathcal{C}$. Furthermore we assume that repeated indices are summed over. In momentum space a finite cut-off $\Lambda$ is applied to regularize eventual UV divergencies. Consequently, we are effectively discussing a lattice system with a finite quasi-particle bandwidth. The bare mass $m^2(t) =  m_0^2 - A \cos(\Omega t)$ is driven with amplitude $A$ and frequency $\Omega$, which, in a linear response regime ($A\ll\Omega,m_0$) creates pairs of excitations.  

It is convenient to rescale time $t\rightarrow 2t/\Omega$ and the fields $\Phi_a \rightarrow (2/\Omega)^{1/2} \Phi_a$, and to introduce the effective coupling constants in the presence of an external drive:
\begin{equation}
g = \frac{2A}{\Omega^2},\qquad u= \frac{8\lambda}{\Omega^3}\;.
\label{eq:couplings}
\end{equation}
The driving amplitude is rescaled by $\Omega^2$, which is a consequence of the relativistic form of the action. The model \eqref{eq:classicalS} displays an equilibrium phase transition to an ordered, symmetry broken phase for $m_0^2<0$ and small $\lambda<\lambda_c$ at low temperatures. The drive destroys the ordered phase already at leading order in $N$~\cite{Chandran2016a}. Hence, as we are interested in the long-time dynamics, we restrict ourselves to initial states in the symmetric phase. Furthermore, in the case of symmetric initial states, we find the same qualitative behavior in all spatial dimensions $d=1, 2, 3$, and thus the presented results focus on $d=1$. We emphasize that our results represent the thermodynamic limit, and thus should be contrasted to the exact diagonalization of small systems.

\textit{\textbf{Nonequilibrium Keldysh formalism.---}}In order to simulate the dynamics of the driven system, we use the nonequilibrium Keldysh formalism~\cite{Keldysh1965}. The time evolution of the two-point contour ordered Green's function $\hat{G}$ is governed by the self-consistent Dyson equation 
\begin{align}
\left(\Box_{t, x} + m^2(t)\right) \hat{G}(t, t')  &+ i \int_\mathcal{C}dt'' \hat{\Sigma}(t, t'') \ast \hat{G}(t'', t')\notag \\
&= -i \hat{\mathbbm{1}} \delta_\mathcal{C}(t-t'),
\label{eq:Dyson}
\end{align}
where $\Box_{t, x}$ is the d'Alembert operator and the self-energy $\hat{\Sigma}$ is given as the functional derivative of the 2PI effective action $\Gamma_2$~\cite{Cornwall1974}, see supplementary material for details~\cite{supp}. The advantages of this approach are that it operates in the thermodynamic limit and respects the conservation laws associated with the global symmetries of the microscopic action, such as energy or momentum conservation. We decompose the contour ordered Green's function as $\hat{G}(t, t') = \hat{F}(t, t') -i/2 \mathrm{sgn}_\mathcal{C}(t - t') \hat{\rho}(t, t')$, with the Keldysh or statistical correlation function $\hat F$, that is symmetric under a permutation of arguments, and the spectral function $\hat \rho$, that is antisymmetric when permuting the arguments.

We employ a $1/N$ fluctuation expansion to the real-time effective action $\Gamma_2$ to next-to-leading order (NLO)~\cite{Berges2002a, Aarts2002}. While in the symmetric phase only a single diagram contributes at leading-order (LO), at NLO an infinite series of diagrams has to be summed. The self-energy up to NLO can be schematically represented by the following diagrammatic series
\begin{align}
&\hspace{-5pt}\Sigma^\text{NLO} = \hspace{-5pt} \parbox{40pt}{\includegraphics[scale=0.45]{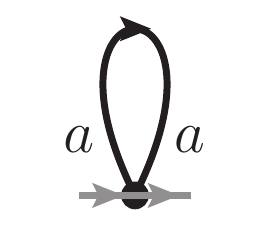}} \hspace{-12pt} +  \hspace{-3pt}\parbox{40pt}{\includegraphics[scale=0.45]{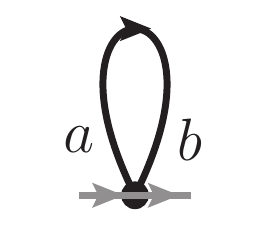}}  \hspace{-12pt} +  \hspace{-3pt}\parbox{40pt}{\includegraphics[scale=0.45]{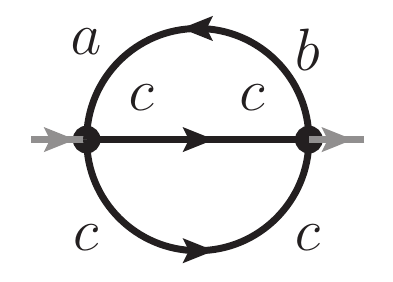}}  \hspace{8pt} +  \hspace{-3pt}\parbox{50pt}{\includegraphics[scale=0.45]{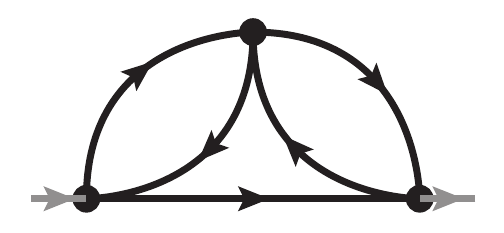}} \hspace{3pt}+\ldots
 \label{eq:selfen}
\end{align}
where lines represent full Green's functions $G$ and dots vertices, each of which comes with a factor $\sim \lambda/N$. In this scheme, the LO (first diagram) is equivalent to a self-consistent Hartree-Fock approximation and thus results in a time-local self-energy that solely renormalizes the bare mass~\cite{supp}. A LO analysis is thus not sufficient to answer the question of whether a prethermal state can be stabilized, as it eliminates the possibility of infinite heating from the beginning. Only at NLO [all other diagrams in \eq{eq:selfen}] the self-energy contains parts which are non-local in time and lead to scattering and memory effects that ultimately enable thermalization. 

The NLO evolution equations are integrated numerically for times up to $3.18 \cdot 10^4$ driving cycles. The momentum cutoff is set to $\Lambda = \pi$. As initial condition we use the LO groundstate of the $O(N)$-model for given interaction $u$ and fixed renormalized mass $m^2_\mathrm{eff}=1$, i.e. the bare mass $m^2_0$ gets adjusted accordingly. We have chosen this convention, since the physically relevant observable quantity is the renormalized mass $m^2_\mathrm{eff}$, which has to be fixed to get comparable results. Furthermore, we set the drive amplitude to $g=1/4$ and scan the interaction $u$ and drive frequency $\Omega$.

\textit{\textbf{Dynamics of the energy density.---}}The central observable to study heating in any driven system is the energy density $\epsilon(t) = \langle\hat{H}(t)\rangle/V$, where $V$ is the system volume. In our scheme the expectation value of the Hamiltonian is directly available from the Keldysh Green's function $F$. Calculating the expectation of the quadratic part of the Hamiltonian is straightforward, whereas for the quartic term, we use Heisenberg's equations of motion to express it in terms of higher order time derivatives of the Keldysh Green's function. We obtain
\begin{equation}
\epsilon(t) = \frac{N}{4} \int_p{\left(2\partial_t \partial_t' - \partial_t^2 + \frac{4\left(p^2 + m_0^2\right)}{\Omega^2}\right)F(t, t', p)}|_{t=t'}.
\label{eq:Energy}
\end{equation} 
Typical plots of $\epsilon(t)$ are shown in Fig.~\ref{fig:timetraces}. We can divide the heating of the system into three regimes: (I) At short times, up to the interaction timescale $t_\mathrm{int}$, the dynamics is dominated by single-particle rearrangements, leading to exponentially fast heating. In that regime, a LO approximation is sufficient to describe the dynamics and scattering of quasi-particles is essentially irrelevant. We define, the interaction timescale $t_\mathrm{int}$ as the time at which the LO and NLO results starts to deviate, which characterizes the time at which non-local contributions to the self-energy become important. 
(II) After this initial stage of heating, the system quickly enters a prethermal plateau with low absorption. This Floquet prethermal state persists up to the thermalization time $t_\mathrm{th}$ and can span several decades in time, thus providing a solid regime for Floquet engineering.
(III) At late times $t \gtrsim t_\mathrm{th}$ heating becomes significant and we expect the system to approach the infinite temperature state. In that regime our data suggests a power-law growth of the energy density $\epsilon(t)\sim t^{\alpha}$. In the following, we discuss these regimes and where possible provide analytical arguments for the observed behavior.

\textit{\textbf{Short time dynamics.---}}At short times, NLO corrections are essentially irrelevant for the dynamics, as confirmed explicitly by comparing LO and NLO results; inset in \figc{fig:tth}{a}. At LO, the system is equivalent to a multi-dimensional anharmonic oscillator with periodically modulated frequencies (see supplementary material~\cite{supp}). We can understand the dynamics in terms of a parametric resonance with the resonance condition set by $\Omega = 2 \omega(p^\star)$, where $\omega(p)=\sqrt{p^2+m_\mathrm{eff}^2}$ is the initial dispersion relation of excitations. The momentum-mode $p^\star$ grows exponentially  and the fastest growing observable will be $F(t, t, p^\star)\sim e^{2 \gamma_{p^\star} t}$. Consequently, using~\eqref{eq:Energy}, the energy density will also grow exponentially in time $\epsilon(t)\sim e^{2\gamma_{p^\star} t}$. In the Gaussian limit, $u=0$, this exponential growth would last indefinitely, but for finite $u$ the self-consistently determined effective mass grows simultaneously with $F(t, t, p^\star)$, breaking the resonance condition at a certain time, and preventing any further energy-absorption in a LO approximation~\cite{Chandran2016a}.

Taking into account NLO corrections quasi-particle excitations interact with each other which will eventually lead to heating. We estimate the validity of the LO calculation by $F(t, t, p^\star) \sim u^{-2/3}$, which determines the time when the first non-trivial diagrammatic contribution [sunset diagram, i.e., third diagram in \eq{eq:selfen}], becomes relevant~\cite{Berges2002,supp}. Considering the exponential growth of $F(t, t, p^\star)$, the interaction timescale obeys the scaling $t_\mathrm{int} \sim (2\gamma_{p^\star})^{-1} \log u^{-2/3}$. The logarithmic scaling of $t_\mathrm{int}$ with $u$ is confirmed in \figc{fig:tth}{a}. Deviations from the logarithmic scaling exist for $u\gtrsim 1$, as in the strong interaction regime NLO processes are important already at initial times, which renders the interaction time scale ill-defined.

In order to validate that the scattering of quasi-particle excitations is the reason for the deviation of the LO and NLO results, we derive a Floquet Fermi's Golden rule (FFGR)~\cite{Knap2015}, which formally considers NLO diagrams with the lowest number of interaction vertices (sunset diagram)~\cite{supp}. We find perfect agreement between the interaction time $t_\mathrm{int}$ evaluated with the full NLO calculation and the FFGR, respectively, which demonstrates that scattering of created excitations is responsible for the deviations between the leading and next-to-leading order time evolution. This explains why the system can heat up further: Once scattering is taken into account, not only pairs of quasi-particles can be created but the energy can also be distributed over many excitations.

\begin{figure}
\includegraphics[width=.48\textwidth]{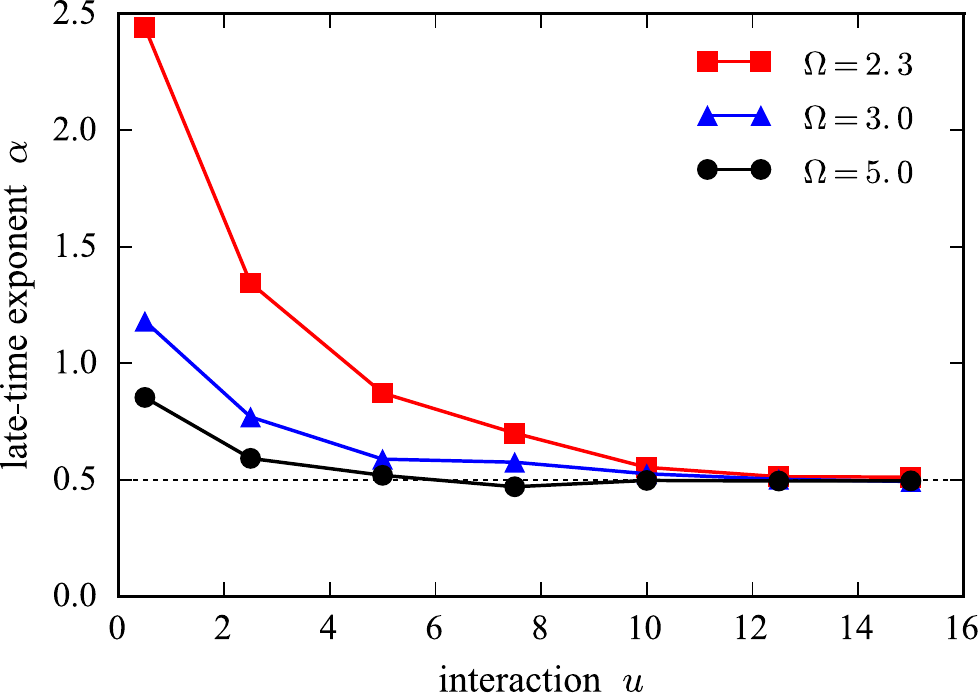}
\caption{\textbf{Powerlaw exponent $\alpha$ of the absorbed energy in the thermalizing regime.} The exponent $\alpha$ is extracted from the algebraic growth of the energy density at late times $\epsilon(t)\sim t^\alpha$. With increasing interaction strength $u$ and driving frequency $\Omega$ (but still within the single-particle band), the exponent $\alpha$ quickly approaches $1/2$, suggesting that in the asymptotic long-time limit, $t\to \infty$, the heating rate universally scales as $\epsilon(t) \sim \sqrt{t}$. }
\label{fig:exponent}
\end{figure}

\textit{\textbf{Floquet prethermalization.---}}Once the parametric resonance regime is left, heating becomes extremely slow and the prethermal plateau is entered. In that regime the number of quasi-particles is small and hence the multi-particle scattering, which is enabling further energy absorption, is much slower than pair creation.
The number of quasi-particle excitations is directly related to the equal time Keldysh Green's function $F$, which due to the self-consistent feedback continues to grow. As the thermalization timescale $t_\mathrm{th}$ is reached, the higher order loop diagrams [\eq{eq:selfen}] that allow for multi-particle scattering start to dominate. Thus, heating becomes significant and the Floquet prethermal state breaks down.

To quantitatively understand the thermalization time scale $t_\mathrm{th}$, we study it as a function of the interaction strength $u$ and driving frequency $\Omega$; \figc{fig:tth}{b}. The thermalization timescale and thus the lifetime of the prethermal plateau decreases with increasing $u$. The dependence is quite strong, with $t_\mathrm{th}$ changing over one order of magnitude as $u$ varies in the interval $[0.5, 15]$ and $\Omega = 2.3$. Fixing the interaction $u$, we find that $t_\mathrm{th}$ decreases with increasing $\Omega$. This is a consequence of all chosen frequencies lying within the initial bandwidth of quasi-particle pairs, $2<\Omega<2\sqrt{\Lambda^2 + 1}$. With increasing $\Omega$, more momentum modes participate in the parametric resonance and consequently the Keldysh component $F$ already ends up being larger as $t_\mathrm{int}$ is reached~\cite{supp}. Based on our previous arguments on the quasi-particle density, the system thus will be earlier driven out of the prethermal plateau.

Even though heating is slow within the prethermal regime $t_\mathrm{int} < t < t_\mathrm{th}$ it remains finite and the system does not become fully stationary. Nevertheless, in this regime the Green's function only depends extremely weakly on the stroboscopic center-of-mass time $T_n = (t+t')/2 = 2\pi n/\Omega$, where $n$ is an integer. Thus, this extremely slow center-of-mass time dependence should not affect the much faster microscopic processes, that are required to realize novel prethermal states.

\textit{\textbf{Thermalization.---}}At times $t\gtrsim t_\mathrm{th}$, the system is driven out of the prethermal regime and the absorption increases. Our numerical simulations suggest that the energy density grows as a powerlaw $\epsilon(t) \sim t^\alpha$ (\fig{fig:timetraces}), which can persist for several decades. We show the exponent $\alpha$ as a function of the interaction strength $u$ for different driving frequencies $\Omega$ in \fig{fig:exponent}. With increasing interaction $u$ and drive frequency $\Omega$ the exponent approaches $1/2$, which appears as a lower bound. In the limit of large $u$ and $\Omega$, the thermalization time scale is smallest and hence, given the fixed maximum time that we can reach in our simulations, the accessible thermalization regime is largest for these parameters. This suggests that the powerlaw exponent might slowly creep to the universal value $1/2$ for any interaction $u$ and drive frequency $\Omega$ in the asymptotic limit, $t\to \infty$. In contrast, we found linear heating at late times in the $O(N)$-model subject to colored noise; see supplemental material~\cite{supp}. Moreover, our results suggest that the driven $O(N)$-model heats to infinite temperature following the well defined prethermal plateau.

\textit{\textbf{Conclusions and Outlook.---}}Our results demonstrate, that a prethermal Floquet state can be stabilized in a generic periodically-driven quantum system, despite strong interactions and despite the driving frequency being comparable to microscopic energy scales of the system. This opens the possibility of realizing exotic states in the Floquet prethermal regime, such as time crystals or other novel symmetry protected topologically phases. Furthermore, our study suggests a universal algebraic heating at late times of the form $\epsilon(t)\sim \sqrt{t}$, which is significantly slower than the linear Joule heating. We attribute this peculiar form of heating to the strong interactions between the dynamically generated quasi particles. How such a sublinear growth can be reconciled with the eigenstate thermalization hypothesis is an important open question. A future study based on Floquet Boltzmann type approach might provide further insights into this behavior.

\begin{acknowledgments}
\textit{\textbf{Acknowledgments.---}}We thank E. Berg, A. Chandran, S. Diehl, A. Silva, and W. Zwerger for interesting discussions. We acknowledge support from the Technical University of Munich - Institute for Advanced Study, funded by the German Excellence Initiative and the European Union FP7 under grant agreement 291763, and from the DFG grant No. KN 1254/1-1.
\end{acknowledgments}

\bibliography{DrivenONPaper_arXiv.bib}

\newpage
\clearpage
\appendix

\setcounter{figure}{0}
\setcounter{equation}{0}

\renewcommand{\thepage}{S\arabic{page}} 
\renewcommand{\thesection}{S\arabic{section}} 
\renewcommand{\thetable}{S\arabic{table}}  
\renewcommand{\thefigure}{S\arabic{figure}} 
\renewcommand{\theequation}{S\arabic{equation}} 

\onecolumngrid

\begin{center}
\textbf{\Large{\large{Supplemental Material: \\Floquet prethermalization and regimes of heating in a periodically driven, interacting quantum system}}}
\end{center}


\subsection{Two-Particle Irreducible Effective Action Approach}

The effective action $\Gamma[\phi, G]$ is the Legendre transform of the generating functional for connected Green's functions $G_{ab}$ and the vacuum expectation value (VEV) $\phi_a$. It can be generally written as~\cite{Berges2002a}
\begin{equation}
\Gamma[\phi, G] = S[\phi] + \frac{i}{2}\rm{tr}_\mathcal{C} \log G^{-1} +\frac{i}{2} \rm{tr}_\mathcal{C} G_0^{-1}G + \Gamma_2[\phi, G].
\label{eq:gam}
\end{equation}
Since we study the system in the symmetric phase, the VEV vanishes and the effective action becomes a functional of $G_{ab}$ only. In \eq{eq:gam}, the free propagator in the symmetric phase is $iG_{0, ab}^{-1} (t, x, t', y) = - ( \Box_{t, x} + m^2 ) \delta_{ab}\delta_\mathcal{C}(t-t')\delta(x-y)$. The functional $\Gamma_2[\phi, G]$ is the two-particle irreducible (2PI) effective action, which is given by the sum of all 2PI vacuum diagrams of 
\begin{equation}
S_\text{int}[\Phi] = -\int_{x, \mathcal{C}} \frac{\lambda}{4!N} \left(\Phi_a(x)\Phi_a(x)\right)^2
\end{equation}
and can be diagrammatically represented as~\cite{Berges2002a}
\begin{align}
\Gamma_{2}[\phi, G] &=\parbox{50pt}{\includegraphics[scale=0.9]{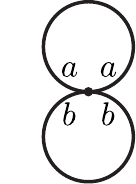}} \hspace{-10pt} +\hspace{-20pt}\parbox{50pt}{\includegraphics[scale=0.9]{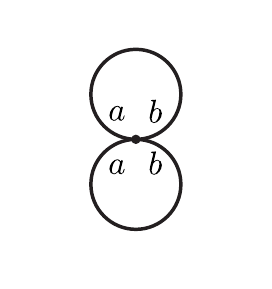}} +\frac{1}{2}\hspace{-5pt}\parbox{65pt}{\includegraphics[scale=0.9]{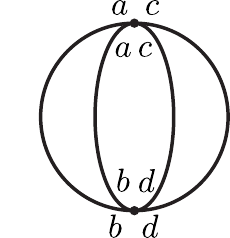}} + \frac{1}{3}\hspace{-3pt}\parbox{66pt}{\includegraphics[scale=0.9]{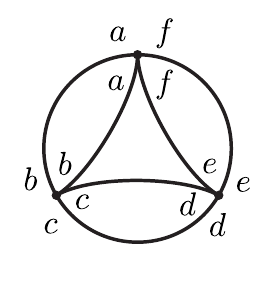}} + \ldots,
\end{align}
where lines denote the propagators $G_{ab}$ and the dots represent interaction vertices $\sim \lambda/N$. The evolution equations of the Green's function, Eq.~\eqref{eq:Dyson}, are obtained from the stationarity condition $\delta \Gamma [G]/\delta G=0$ using the definition of the self-energy $\Sigma_{ab} =2i~ \delta \Gamma_2[G]/\delta G_{ab}$.

The integration over the Keldysh contour $\mathcal{C}$ in Eq.~\eqref{eq:Dyson} can be resolved by parametrizing the contour ordered Green's function $G_{ab}$ in terms of the statistical propagator $F$ and the spectral-function $\rho$ as follows
\begin{align}
G_{ab}(t, x, t', y) &= F_{ab}(t, x, t', y) -\frac{i}{2}\mathrm{sgn}_\mathcal{C}(t-t') \rho_{ab}(t, x, t', y)\notag \\
F_{ab}(t, x, t', y) &= \frac{1}{2}\langle \left\{\Phi_a(t, x), \Phi_b(t', y)\right\}\rangle \notag \\
\rho_{ab}(t, x, t', y) &=  i \langle \left[\Phi_a(t', x), \Phi_b(t, y)\right] \rangle .
\end{align}
Using this parametrization, the causal Kadanoff-Baym equations follow directly from Eq.~\eqref{eq:Dyson}. We further make use of the fact, that the driving is uniform in space and that it conserves the $O(N)$-symmetry, i.e., $G_{ab} = G \delta_{ab}$ to obtain
\begin{align}
&\left[\partial_t^2 + p^2 + m^2(t)  + \Sigma^{(0)}(t)\right] F(t,  t', p) = - \int\limits_0^{t} dt''  \bar\Sigma^\rho (t, t'', p) F(t'', t', p) + \int\limits_0^{t'} dt'' \bar\Sigma^F (t, t'', p) \rho(t'',  t', p) \notag \\
&\left[\partial_t^2 + p^2 + m^2(t) +  \bar\Sigma^{(0)}(t)\right] \rho(t, t', p) = - \int\limits_{t'}^{t} dt''  \bar\Sigma^\rho(t, t'', p) \rho(t'', t', p).
\label{eq:KadanoffBaym}
\end{align}

The self-energy is decomposed as $\Sigma(t, t', p) = -i \Sigma^{(0)}(t) \delta_\mathcal{C}(t-t') + \bar{\Sigma}(t, t', p)$. Here, $\Sigma^{(0)}(t)$, is the local contribution to the self-energy, which leads to a mass renormalization $m_\mathrm{eff}^2 = m^2 + \Sigma^{(0)}$. By contrast,  $\bar{\Sigma}(t, t', p)$, is nonlocal in time and splits into spectral and statistical components, analogously to the Green's function.

To make use of Eq.~\eqref{eq:KadanoffBaym}, we need the 2PI effective action $\Gamma_2[G]$. As the exact form of $\Gamma_2[G]$ is unknown for our interacting model, we employ a large-$N$ approximation scheme~\cite{Berges2002a, Aarts2002}. At next-to-leading order (NLO), the self-energy becomes
\begin{align}
\Sigma^{F, \text{NLO}}(t, x, t', y) &= -\frac{\lambda}{3N}\left(F(t, x, t', y)I_F(t, x, t', y) - \frac{1}{4}\rho(t, x, t', y)I_\rho(t, x, t', y)\right)\notag\\
\Sigma^{\rho, \text{NLO}}(t, x, t', y) &= -\frac{\lambda}{3N}\left(F(t, x, t', y)I_\rho(t, x, t', y) + \rho(t, x, t', y)I_F(t, x, t', y)\right)\notag\\
\Sigma^{(0)}(t) &= \lambda \frac{N+2}{6N} F(t, x, t, x).
\label{eq:SigmaNLO}
\end{align}

The functions $I_F$, $I_\rho$ are often referred to as summation functions and obey the integral equations
\begin{align}
I_F(t, x, t', y) &= \Pi_F(t, x, t', y) - \int\limits_0^{t}dt'' I_\rho(t, x, t'', y) \Pi_F(t'', x, t', y) + \int\limits_0^{t'} dt'' I_F(t, x, t'', y) \Pi_\rho(t'', x, t', y) \notag\\
I_\rho(t, x, t', y) &= \Pi_\rho(t, x, t', y)- \int\limits_{t'}^{x^0}dt'' I_\rho(t, x, t'', y) \Pi_\rho(t'', x, t', y),
\label{eq:SumFunc}
\end{align}
with the polarization bubble $\Pi(t, x, t', y) = \frac{\lambda}{6} G(t, x, t', y) G(t, x, t', y)$.

The set of Equations (\ref{eq:KadanoffBaym}--\ref{eq:SumFunc}) have to be solved simultaneously, starting from $t=t'=0$. To this end, we discretize the system in momentum space and sample $46$ points, which we have checked to be large enough to describe the thermodynamic limit. The equations of motion are then integrated numerically using a leap-frog method. 

\subsection{Dynamics at Leading Order}

\begin{figure}
\includegraphics{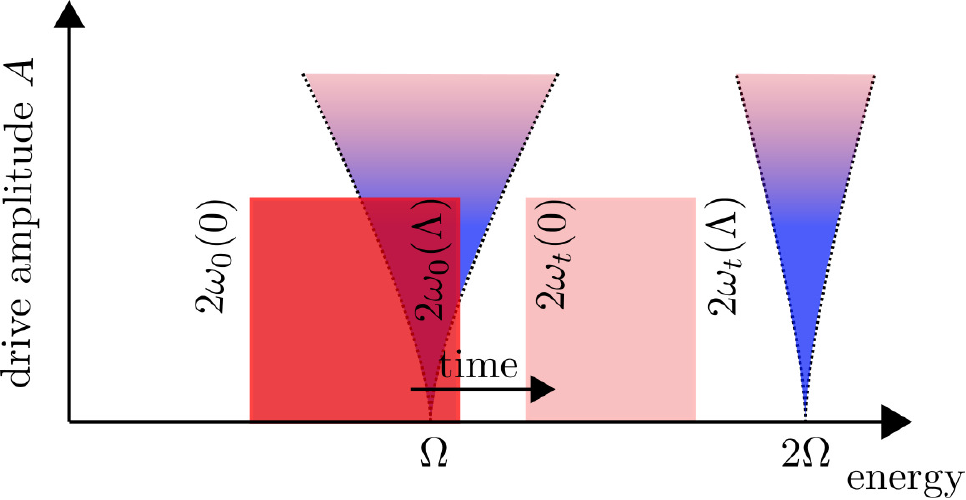}
\caption{\textbf{Illustration of the parametric resonance.} When a part of the initial bandwidth (red shaded region) for quasi-particle pairs lies in a parametric resonance (blue shaded regions), the corresponding modes grow exponentially, which leads to an exponentially growth of the effective mass $m_\mathrm{eff}(t)$. Accordingly the dispersion relation $\omega_t(p)$ is shifted toward higher energies (light red shaded region). Once the quasi-particle bandwidth lies completely in between to region of parametric resonance, the system stops absorbing energy in the leading order approximation. The parametric resonance at $n\Omega$, $n=1, 2, \dots$ is only sharp in the limit of vanishing drive amplitude $A$ and smears out with increasing $A$. For small $A$, the width of the resonance grows linearly with $A$.}
\label{fig:ParamRes}
\end{figure}

To leading order the self-energy is time-local and the evolution equations simplify to
\begin{align}
&\left[\partial_t^2 + p^2 + m^2_0 + A \cos\Omega t + \frac{\lambda}{6} \int_q F(t, t, q)\right] F(t,  t', p) = 0 \notag \\
&\left[\partial_t^2 + p^2 + m^2_0 +A \cos\Omega t +  \frac{\lambda}{6} \int_q F(t, t, q)\right] \rho(t, t', p) =0
\label{eq:KadanoffBaymLO}
\end{align}

The equations \eqref{eq:KadanoffBaymLO} describe coupled anharmonic parametric oscillators (one oscillator for each $t'$ and $p$) with initial eigenfrequencies $\omega_0(p) = \sqrt{p^2+m_\mathrm{eff}(0)^2}$. Let us first discuss the entirely noninteracting case $\lambda=0$, in which Eqs.~\eqref{eq:KadanoffBaymLO} are independent Mathieu equations. It is known from classical mechanics, that the modes satisfying the resonance condition $2\omega_0(p) = n\Omega$ with $n=1, 2, \dots$ experience a parametric resonance and will grow exponentially in time. As there is no feedback on the spectrum of the system for $\lambda = 0$ this exponential growth in the resonant modes continues forever.

For finite $\lambda$, the exponential growth of the statistical correlation function $F(t, t', p)$ for momenta $p$ satisfying the resonance condition leads to an exponential growth of the effective mass $m_\mathrm{eff}^2(t) = m^2(t) + \lambda/6 \int_q F(t, t, q)$, which shifts the dispersion of quasi-particles to higher energies and reduces the effective quasiparticle bandwidth $2[\sqrt{\Lambda^2+m_\text{eff}(t)^2}-m_\text{eff}(t)]$~\cite{Chandran2016a}. Therefore, the quasi-particle bandwidth will at a certain time lie entirely in between the parametric resonances and the system cannot absorb energy anymore, \fig{fig:ParamRes}. 

The failure of the system to absorb further energy can be traced back to the fact, that the LO self-energy is local in time and only leads to a renormalization of the quasi-particle dispersion. Except for this renormalization the quasi-particles remain sharp excitations and there is no mechanism present, that allows energy-transfer between them. Consequently, there is only energy absorption from the drive when the driving frequency hits the sharp resonance for the creation of quasi-particle pairs and the heating stops as soon as the resonance condition cannot be fulfilled anymore.

\subsection{Floquet Fermi's Golden Rule}

\begin{figure}
\includegraphics[scale=0.7]{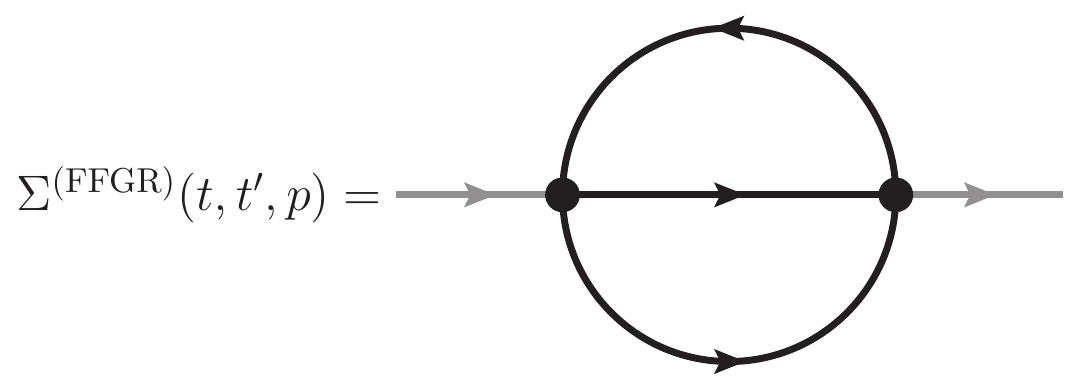}
\caption{\textbf{Floquet Fermi's golden rule.} The "sunset" diagram takes into account scattering of two quasi-particles beyond a simple renormalization of the quasi-particle mass. It is non-local in time and leads to memory effects.}
\label{fig:sunset}
\end{figure}

The simplest diagram leading to scattering between quasi-particles is the "sunset" diagram, see Fig.~\ref{fig:sunset}. This diagrams includes interactions of only two quasi-particles. By contrast, higher loop diagrams would include scattering events of more than two particles. As we discuss in the main text, these higher-order events become relevant only at later times.

We obtain the following expression for the FFGR self-energy
\begin{align}
\Sigma^{(\mathrm{FFGR})} (t, t', p) = - \frac{\lambda}{18N} \int\limits_{k, q} G(t, t', p+q) G(t, t', k) G(t', t, q+l).
\label{eq:FFGR1}
\end{align}
Splitting Eq.~\eqref{eq:FFGR1} into statistical and spectral components, we obtain 
\begin{align}
&\Sigma^{F, \text{FFGR}}(t, t', p) = -\frac{\lambda}{3N}\int\limits_q\left(F(t, t', p+q)\Pi_F(t, t', q) - \frac{1}{4}\rho(t, t', p+q)\Pi_\rho(t, t', q)\right)\notag\\
&\Sigma^{\rho, \text{FFGR}}(t, t', p) = -\frac{\lambda}{3N}\int\limits_q\left(F(t, t', p+q)\Pi_\rho(t, t', q) + \rho(t, t', p+q)\Pi_F(t, t', q)\right).
\label{eq:FFGR2}
\end{align}
Expressing $\Sigma^{\mathrm{FFGR}}$ in this way, we see that the Floquet Fermi's golden rule analysis amounts to replacing the summation function $I$ in the expression for the NLO self-energy, Eq.~\eqref{eq:SigmaNLO}, with the polarization bubble $\Pi$.

We calculate the energy-density of the system resulting from the FFGR and find very good agreement with the NLO results for $t_\mathrm{int}$, see Fig.~\ref{fig:tth} (a). However, we emphasize that the FFGR self-energy, Eq.~\eqref{eq:FFGR2}, does not correspond to a conserving expansion of the 2PI effective action $\Gamma_2$ and hence is bound to fail for long times, as it eventually becomes divergent. 

\subsection{Multiplicative Noise}

We study the leading-order time evolution of the statistical Green's function subject to multiplicative noise
\begin{equation}
\left[\partial_t^2 + p^2 + m^2_0 + A \cos\Omega t + \frac{\lambda}{6} \int_q F(t, t, q) + \xi(t)\right] F(t,  t', p) = 0.
\label{eq:LONoise}
\end{equation}
Introducing noise $\xi(t)$ is expected to mimic, at least very crudely, the effect of scattering. Therefore, the system is expected to heat to infinite temperature even with the leading order self-energy. 

We explore two cases for the random process, which are white noise and correlated noise, respectively. In the case of white noise, $\xi_\mathrm{w}(t)$  reduce to Gaussian random variables with vanishing mean, $\langle\xi_\mathrm{w}(t)\rangle = 0$ and auto-correlation $\langle\xi_\mathrm{w}(t)\xi_\mathrm{w}(t')\rangle = \gamma^2 \delta(t-t')$. By contrast, the correlated noise $\xi_\mathrm{c}(t)$ obeys the stochastic differential equation of the Ornstein-Uhlenbeck process
\begin{equation}
d\xi_\mathrm{c}(t) = -\frac{1}{\tau} \xi_\mathrm{c}(t) dt + \sigma dW(t),~\xi_\mathrm{c}(0) = 0,
\label{eq:OUprocess}
\end{equation}
where $\tau$ is the correlation time, $\sigma$ controls the strength of the noise, and $W(t)$ is the standard Brownian motion. The auto-correlation of $\xi_\mathrm{c}$ is given by
\begin{equation}
\langle \xi_\mathrm{c}(t)\xi_\mathrm{c}(t')\rangle = \frac{\sigma^2\tau}{2}\left(e^{-\frac{|t-t'|}{\tau}} - e^{-\frac{t+t'}{\tau}}\right)
\label{eq:OUcorr}
\end{equation}
and $\langle\xi_\mathrm{c}(t)\rangle=0$. Note that white noise is recovered in the limit $\tau \rightarrow 0$, $\sigma \rightarrow \infty$, keeping $\sigma \tau = \gamma$ fixed.

\begin{figure}
\includegraphics{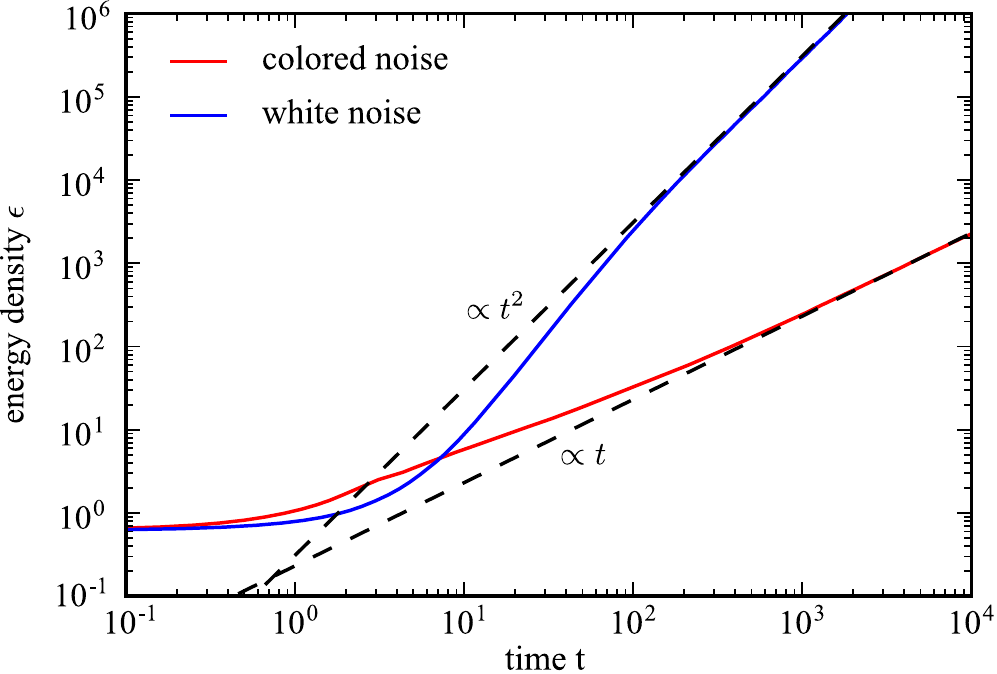}
\caption{\textbf{Time evolution of the energy density obtained from leading order equations subject to noise.} With multiplicative noise the system absorbs energy indefinitely even at leading order. The energy density grows with a powerlaw $\epsilon(t)\sim t^\alpha$ for late times for arbitrary system parameters. Depending on whether the noise is white, i.e., completely uncorrelated in time or colored, i.e., correlated in time, the growth is quadratic or linear, respectively. From that we deduce that correlations in time slow down heating in the system. The data is shown for driving frequency $\Omega = 2.3$ and interaction strength $u=1.0$. The strength of the white noise is $\gamma = 2.0/\Omega^2$ whereas for colored noise we have chosen $\sigma = 2/\Omega^2$ and $\tau=20/\Omega$.}
\label{fig:noise}
\end{figure}

White noise is completely uncorrelated, while colored noise has exponentially decaying correlations in time. Hence, one expects that the system thermalizes faster when it is subject to white noise. This is what we find by numerically solving \eq{eq:KadanoffBaymLO}. Moreover, we find that the energy-density grows according to a power-law $\epsilon(t) \sim t^\alpha$; \fig{fig:noise}. We exploit the similarity of Eq.~\eqref{eq:KadanoffBaymLO} and an anharmonic oscillator, for which it has been shown that the energy grows quadratically in time for white noise ($\alpha=2$), whereas colored noise leads to a linear growth $\epsilon \sim t$ ($\alpha=1$)~\cite{MallickMarcq2002, MallickMarcq2003}. The dynamical evolution in our system, \eq{eq:LONoise}, confirms these expectations. Therefore, the heating due to either white ($\epsilon\sim t^2$) or colored ($\epsilon\sim t$) noise is substantially faster than the asymptotic heating we observe when solving the equations of motion self-consistently up to NLO ($\epsilon \sim \sqrt{t}$). We attribute the slow heating obtained from the full solution up to NLO to the strong interactions between quasi-particles which cannot be simply mimicked by multiplicative noise.

\end{document}